\documentclass[12pt,epsfig]{article}
\usepackage{graphicx,amsmath,amssymb}

\newlength{\dinwidth}
\newlength{\dinmargin}
\def\lapproxeq{\lower .7ex\hbox{$\;\stackrel{\textstyle
<}{\sim}\;$}}
\def\gapproxeq{\lower .7ex\hbox{$\;\stackrel{\textstyle
>}{\sim}\;$}}
\def\gtrsim{\lower .7ex\hbox{$\;\stackrel{\textstyle
>}{\sim}\;$}}
\def\lesim{\lower .7ex\hbox{$\;\stackrel{\textstyle
<}{\sim}\;$}}

\def\be{\begin{equation}}
\def\ee{\end{equation}}
\def\bea{\begin{eqnarray}}
\def\eea{\end{eqnarray}}

\def\pp{p\bar{p}}
\def\ra{ \rightarrow }

\begin{document}
\begin{flushright}
IPPP/06/11 \\
DCPT/06/22 \\
24th February 2006 \\

\end{flushright}

\vspace*{0.5cm}

\begin{center}
{\Large \bf On the role of hard rescattering in exclusive diffractive Higgs production}

\vspace*{1cm}
\textsc{V.A.~Khoze$^{a,b}$, A.D. Martin$^a$ and M.G. Ryskin$^{a,b}$} \\

\vspace*{0.5cm}
$^a$ Department of Physics and Institute for
Particle Physics Phenomenology, \\
University of Durham, DH1 3LE, UK \\
$^b$ Petersburg Nuclear Physics Institute, Gatchina,
St.~Petersburg, 188300, Russia \\

\end{center}

\vspace*{0.5cm}

\begin{abstract}
We discuss the contribution of so-called semi-enhanced hard rescattering corrections to central exclusive diffractive
Higgs production, $pp \ra p+H+p$, at the LHC. We present arguments to show that these corrections are
small.  We confirm these expectations by considering HERA data for leading neutron production.
\end{abstract}

\section{Introduction}
Central exclusive diffractive processes offer an excellent opportunity to study the Higgs sector at the LHC in an exceptionally clean environment; for recent reviews see, for example, \cite{rev}. The process we have in mind is
\begin{equation}
 \label{excl}
 pp\to p\; +\; H\; +\; p
\end{equation}
where the + signs denote large rapidity gaps.   
Demanding such an exclusive process (\ref{excl}) leads to a small cross section \cite{KMR}.
At the LHC, we predict
\begin{equation}
\sigma_{\rm excl}(H)~\sim ~10^{-4}~\sigma^{\rm tot}_{\rm incl}(H).
\end{equation}
In spite of this, the exclusive reaction (\ref{excl}) has the following advantages:
\begin{itemize}
\item[(a)]
The mass of the Higgs boson can be measured
with high accuracy (with mass resolution $\sigma(M)\sim 1$ GeV) by measuring the
missing mass to the forward outgoing protons, {\it provided} that they can be accurately tagged far away from the interaction point. Such a measurement can be done irrespective of the decay mode, and is at the heart of an LHC proposal
\cite{FP420} to complement the central detectors by forward proton taggers in the 420m region from the interaction
point.
\item[(b)]
The leading order $b\bar b$
  QCD background is suppressed by the P-even $J_z=0$ selection
rule \cite{KMRmm}, where the $z$ axis is along the direction of the proton beam.
Therefore one can consider the observation of a Standard Model Higgs boson via $H\to
b\bar b$, which is the main decay mode for a mass $M \lapproxeq 140$ GeV. Moreover, a measurement of the mass of the decay products must match the `missing mass' measurement. It should be possible to achieve a signal-to-background ratio of the order of 1. For an integrated
LHC luminosity of ${\cal L} \sim 60 ~{\rm fb}^{-1}$ we expect about a dozen or so observable events
for a Standard Model Higgs, {\it after} accounting for signal efficiencies and various cuts\footnote{See Ref.~\cite{DKMOR} for early estimates of the signal-to-background ratio.}.
\item[(c)]
The quantum numbers of the central object (in particular, the
C- and P-parities) can be analysed by studying the azimuthal angle
distribution of the tagged protons \cite{Centr}. Due to the selection
rules, the production of $0^{++}$ states is strongly favoured.
\item[(d)]
There is a very clean environment for the
exclusive process -- the soft background is strongly suppressed.
\item[(e)]
Extending the study to SUSY Higgs bosons, there are regions of SUSY parameter space were the
signal is enhanced by a factor of 10 or more, while the background remains unaltered.  Indeed,
there are even regions where the conventional inclusive Higgs search modes are suppressed, whereas the
exclusive diffractive signal is
enhanced, and even such that both the $h$ and $H$ $0^{++}$ bosons may be detected \cite{KKMRext}.   
\end{itemize}

\section{The KMR estimate of $pp \ra p+H+p$ at the LHC}

\begin{figure}
\begin{center}
\includegraphics[height=15cm]{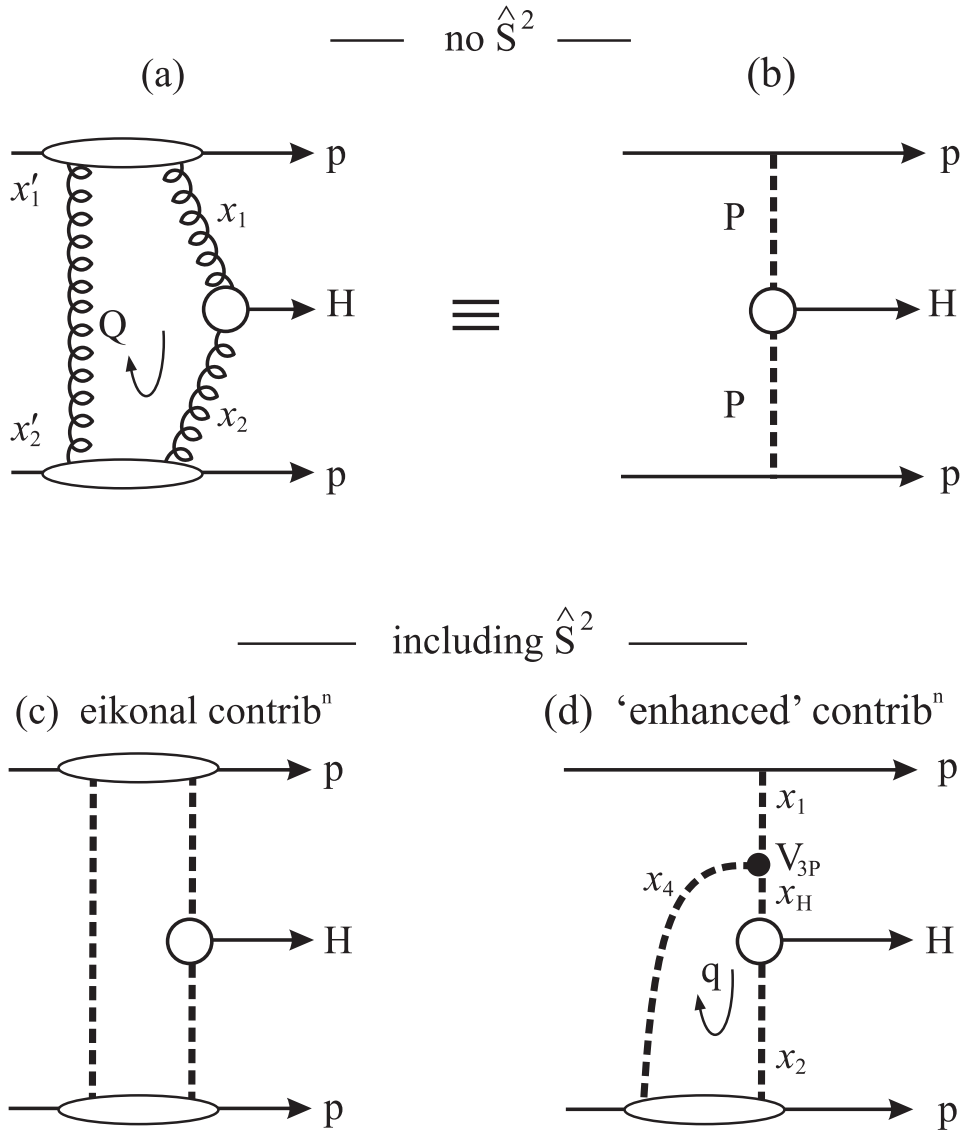}
\caption{Schematic diagrams for central exclusive Higgs production,
$pp \to p+H+p$. The presence of Sudakov form factors ensures the infrared
stability of the $Q_t$ integral over the gluon loop in diagram (a). It is also necessary
to compute the probability, ${\hat S}^2$, that the rapidity gaps survive soft and semi-hard rescattering;
the two possible types of contributions are shown in diagrams (c) and (d) respectively, where the dashed lines
represent Pomeron exchanges (as in version (b) of diagram (a)).  
In addition to diagram (d), there is a `mirror-imaged' enhanced diagram with
the additional Pomeron instead being emitted from the upper proton, and an enhanced diagram with
additional Pomerons being emitted
from both protons and coupling to intermediate partons of the other proton. The expectation is that diagram (c) gives ${\hat S}^2\simeq 0.026$ at the LHC,
whereas in the text we argue that the enhanced diagrams do not give a significant contribution.}
\label{fig:H}
\end{center}
\end{figure}

The basic mechanism for the exclusive process, $pp\ra p+H+p$, is
shown in Fig.~$\ref{fig:H}$(a). The left-hand gluon $Q$ is needed to screen the
colour flow caused by the active gluons labelled by $x_1$ and $x_2$. The $t$-integrated cross section is of the form \cite{KMR,INC}
\begin{equation}
\sigma \sim \frac{{\hat S}^2}{b^2} \left| N\int\frac{dQ^2_t}{Q^4_t}\: f_g(x_1, x_1', Q_t^2, \mu^2)f_g(x_2,x_2',Q_t^2,\mu^2)~ \right| ^2, \label{eq:M}
\end{equation}
where $b/2$ is the $t$-slope of the proton-Pomeron vertex, and the constant $N$ is known in terms of the $H\to gg$ decay width. 
The factor, ${\hat S}^2 $, is the probability that the rapidity gaps survive against population by secondary hadrons.  It has been omitted (${\hat S}^2 =1$) in Fig.~$\ref{fig:H}$(a).  We will consider it in a moment. 
The amplitude-squared factor, $|M_0|^2$, however, may
be calculated using perturbative QCD techniques, since the dominant contribution to the integral comes from the region $\Lambda_{\rm QCD}^2\ll Q_t^2\ll M_H^2$. 
The probability amplitudes, $f_g$, to find the appropriate pairs of
$t$-channel gluons ($x_1,x_1'$) and ($x_2,x_2'$), are given by the skewed
  unintegrated gluon densities at a {\it hard} scale $\mu \sim M_H/2$.

Since the momentum fraction $x'$ transferred through the
screening gluon $Q$ is much smaller than that ($x$) transferred through
the active gluons $(x'\sim Q_t/\sqrt s\ll x\sim M_H/\sqrt s\ll 1)$, it
is possible to express $f_g(x,x',Q_t^2,\mu^2)$
in terms of the conventional integrated density
$g(x)$. A simplified form of this relation is \cite{KMR}
\begin{equation}
\label{eq:a61}
  f_g (x, x^\prime, Q_t^2, \mu^2) \; = \; R_g \:
\frac{\partial}{\partial \ln Q_t^2}\left [ \sqrt{T_g (Q_t, \mu)} \: xg
  (x, Q_t^2) \right ], 
\end{equation} 
which holds to 10--20\%
accuracy.
The factor $R_g$ accounts for
the single $\log Q^2$ skewed effect.  It is found to
be about 1.4 at the Tevatron energy and about 1.2 at the energy of the LHC.

Note that the $f_g$'s embody a Sudakov suppression
factor $T$, which ensures that the gluon does not radiate in the
evolution from $Q_t$ up to the hard scale $\mu \sim M_H/2$, and so
preserves the rapidity gaps. The Sudakov factor is \cite{WMR}
\begin{equation}
\label{eq:a71}
  T_g (Q_t, \mu)=\exp \left (-\int_{Q_t^2}^{\mu^2}
  \frac{\alpha_S (k_t^2)}{2 \pi}\frac{dk_t^2}{k_t^2} \left[
  \int_\Delta^{1-\Delta}zP_{gg} (z)dz
\ + \ \int_0^1 \sum_q\
  P_{qg} (z)dz\right]\right),
\end{equation}
with $\Delta = k_t/(\mu + k_t)$.  The square root arises in
(\ref{eq:a61}) because the (survival) probability not to emit any
additional gluons  is only relevant to
the hard (active) gluon.  It is the presence of this Sudakov factor
which makes the integration in (\ref{eq:M}) infrared stable, and
perturbative QCD applicable.

In fact, the
$T$-factors have been calculated to {\it single} log
accuracy \cite{KKMRext}. The collinear single logarithms may be summed up using the
DGLAP equation. To account for the `soft' logarithms (corresponding
to the emission of low energy gluons) the one-loop virtual correction
to the $gg\to H$ vertex was calculated explicitly, and then the scale
$\mu=0.62\ M_H$ was chosen in such a way that eq.(\ref{eq:a71})
reproduces the result of this explicit calculation. It is sufficient to
calculate just the one-loop correction since it is known that the
effect of `soft' gluon emission exponentiates. Thus
(\ref{eq:a71}) gives the $T$-factor to single log accuracy.

Now we discuss the rapidity gap survival factor, ${\hat S}^2 $.  It has been calculated using an eikonal model which embodies all the main features of soft diffraction.  A schematic diagram is shown in Fig.~$\ref{fig:H}$(c).
The additional Pomeron may couple the upper and lower proton lines in all possible configurations. 
It is found to be ${\hat S}^2 \simeq 0.026$ for $pp\ra p+H+p$ at the LHC.  The uncertainty in the eikonal evaluation
of  ${\hat S}^2 $ has been estimated to be $\pm 50\%$ \cite{KKMRext,KMRsoft}.  In this connection it is interesting to note
that an alternative determination, based on a Monte Carlo calculation, also yields ${\hat S}^2 =0.026$ at the LHC 
\cite{lonnblad}.  A review of the various determinations of ${\hat S}^2$, showing general agreement, can be found in \cite{gotsman}. Moreover the value ${\hat S}^2=0.024$ was found in the recent study described in \cite{BM},
where the amplitude shown in Fig.~$\ref{fig:H}$(a) was denoted TPF (Two Pomeron Fusion).
Actually the exclusive cross section is proportional to
the factor ${\hat S}^2/b^2$, which is almost constant in the relevant interval $b=4~-~6$ GeV$^{-2}$ \cite{Sb}, 
where $b/2$ is the $t$-slope of the proton-Pomeron vertex.

\section{Enhanced diagrams: theoretical uncertainties} 

Besides the uncertainties in the gap survival factor ${\hat S}^2 $
caused by the soft eikonal rescattering of the incoming (outgoing)
protons 
there is the possibility of an additional effect.  The gap may be filled by the
secondaries created in the rescattering of the intermediate
partons; see, for example, \cite{enh}. Formally this effect is described by the semi-enhanced (and/or
enhanced) reggeon diagrams. One such diagram\footnote{The term {\it enhanced diagram} 
originates from Reggeon Field Theory. It means that,
contrary to eikonal rescattering, we have an additional
integration over the rapidity of $V_{3P}$ vertex. This integration
enhances the contribution of the given graph (rather
than the whole amplitude) by an extra logarithm, arising from the available space
in rapidity. Really Fig.~$\ref{fig:H}$(d), with one $V_{3P}$ vertex is called a {\it semi-enhanced diagram}, whereas
an {\it enhanced diagram} contains two $V_{3P}$ vertices and hence two integrations over their rapidities.} is shown schematically 
in Fig.~$\ref{fig:H}$(d). Since the intermediate gluons have a
relatively large transverse momenta, there a possibility that the contribution may be evaluated
within the framework of perturbative QCD. It is proportional to the QCD coupling $\alpha_s$ times the
density of gluons, generated by the lower proton in the rapidity
interval occupied by the intermediate partons of the upper
proton, that is $f_g(x_4,k^2_{t,4},...)$. Here $x_4$ and $k_{t,4}$ are the momentum fraction of the lower
proton and the transverse momentum carried by the $t$-channel gluon in the upper cell of the gluon
ladder corresponding to the additional Pomeron in Fig.~$\ref{fig:H}$(d).  Note that $x_4$ can become
very small, $\sim 10^{-5}$.

The first detailed attempt to calculate such a contribution within the perturbative QCD framework 
has been performed in ref.\cite{BM}.  They evaluated an amplitude
of the form\footnote{Note that our triple-Pomeron vertex $V_{3P}$ is defined slightly differently to that
in ref.\cite{BM}.}
\be
M_1~\sim~\int \frac{dx_4}{x_4} \int\frac{d^2 q_t}{2\pi^2} \int\frac{d^2 k_{t,4}}{k^4_{t,4}}~f_g(x_4,k^2_{t,4},...)
~V_{3P}~M_0,
\label{eq:enh}
\ee
where the unintegrated gluon density $f_g(x_4,...)$ was calculated using 
the Balitsky-Kovchegov equation \cite{BK}, and where the leading log expression
for the QCD triple-Pomeron vertex, $V_{3P}$, was used \cite{BW}. Their result was that the enhanced diagrams give a rather large (negative)
correction to exclusive Higgs boson production at the LHC energy.  That is the exclusive Higgs signal may
be significantly reduced.

However the computation of the enhanced diagrams has, itself, many unresolved uncertainties. 
{\it First}, the next-to-leading logarithmic (NLL) corrections to the triple-Pomeron vertex
are not known at present. To see the possible effect that these could have, we note that in the original Reggeon
phenomenological calculations a ``threshold'' was usually introduced, such that the rapidity interval between two
Reggeon vertices must exceed $\Delta Y = 2\ -\ 3$ \cite{KT,KPTM}.
An analogous effective repulsion
between the two vertices of gluon emission has also been observed in the calculation of the
NLL BFKL corrections \cite{FLC}.  
The NLL correction, $\omega_1$, to the intercept of the Pomeron trajectory is such that
 \begin{equation}
\label{e1}
\omega^{\rm NLL}~=~\omega_0+\omega_1~\simeq~\omega_0(1-6.2\alpha_s),
\end{equation}
where $\omega \equiv \alpha_P(0)-1$, and where $\omega=\omega_0=(N_c\alpha_s/\pi)4\ln 2~$ is the LO BFKL result.
It turns out that the major part of this NLL
correction is of pure kinematical origin \cite{SAl}.
On the other hand, in the presence of the ``threshold''
$\Delta Y$ we have a behaviour exp$(\omega Y) \sim x^{-\omega}$ where the intercept is given by \cite{schmidt}
\begin{equation}
\label{e2}
\omega~=~\omega_0~e^{-\omega\Delta Y}~=~\omega_0(1-\omega_0\Delta Y+...).
\end{equation}
Thus, if we assume that the whole NLL correction is explained by the
$\Delta Y$ threshold, then, on comparing the decrease of the intercept given by (\ref{e1}) and (\ref{e2}), we
obtain the value
\begin{equation}
\label{e3}
\Delta Y =6.2/((4\ln 2) N_c/\pi)\simeq 2.3,
\end{equation}
which is very close to that coming from the original Regge phenomenology.

If, indeed, the NLL correction to the triple-Pomeron vertex has the form
of a $\Delta Y=2.3$ threshold, then it follows that the semi-enhanced correction
will only contribute when the rapidity interval\footnote{$x_H$ is the proton momentum fraction carried by the Higgs
boson.} $\delta y=y_p-y_H=
\ln (1/x_H)$  between the incoming proton and the vertex of Higgs boson
emission becomes larger than $2\Delta Y$; since the interval between
the rapidity of the triple-Pomeron vertex ($y_V$) and the proton,
 and the interval between the triple-Pomeron and Higgs vertices,
both must exceed $\Delta Y$.  That is, we must have
$$y_p-y_V>\Delta Y,~~~~~~{\rm and}~~~~~~y_V-y_H>\Delta Y.$$
If we impose these requirements, then the semi-enhanced correction
(considered in \cite{BM}) will not contribute significantly\footnote{We thank A.B. Kaidalov for emphasizing the crucial
role of this threshold effect, see also \cite{enh}.} to the
central ($y_H=0$) exclusive production of a Higgs boson of
mass $M_H>140$ GeV at the LHC energy $\sqrt{s}=14$ TeV, since the
available rapidity interval $\delta y=\ln (\sqrt{s}/M_H) <4.6$ is less than
$2\Delta Y$.  Even for $M_H=120$ GeV the available phase space is minute.

{\it Secondly}, at the moment there are no experimental data which determine the partons
in the region with $x \lapproxeq 10^{-4}$.
There is a tendency that at low $Q^2 < 2-3$ GeV$^2$ and $x<10^{-3}$ for the
gluon density to start to decrease with $x$ decreasing \cite{CTEQ,MRST}.  Moreover, in some global analyses
the gluon distribution is even negative for $Q^2=2$ GeV$^2$ and $x<3\times
10^{-4}$ \cite{MRST}.  A more detailed discussion of our present knowledge (and uncertainties)
of the low-$x$ parton distributions can be found, for example, in \cite{MR}.

{\it Finally}, we recall that
infrared stability of the calculation of \eqref{eq:enh} is {\it only} provided by
the so-called `saturation momentum' $Q_s(x_4)$, below which the unintegrated gluon
density $f_g$ becomes proportional to $k^2_t$.  That is
$$ f_g(x_4,k^2_{t,4},...) ~ \propto ~ k^2_{t,4}~~~~~~~~~{\rm for}~~k_{t,4}<Q_s(x_4). $$
Indeed, the dimension of the Pomeron loop $\int d^2 q_t$ integration is
compensated by the infrared-type integral $\int d^2 k_{t,4}/k^4_{t,4}$.
Here the infrared divergency is not protected by Sudakov factors,
and the infrared cutoff is provided either by the inverse proton
size or by the saturation momentum $Q_s$.\footnote{When the
essential values of the Pomeron loop momentum $q_t$ (and $k_{t,4}$)
are much smaller than the value of the gluon transverse momenta $Q_T$ in the
loop which contains the Higgs ($gg\to H$) vertex, we can justify the validity
of the same leading order (LO) P-even, $J_z=0$ selection rule as in
the original amplitude, Fig.~$\ref{fig:H}$(a), without the semi-enhanced correction.} The hope
is that at very low values of $x_4$ the momentum $Q_s(x_4)$ is large
enough for perturbative QCD to be applicable.  This is not excluded; however so far there is
no experimental evidence (in the HERA data) to show the explicit
growth of $Q_s(x)$ with decreasing $x$.

  Thus the size of the correction crucially depends on the gluon
density in the saturation (or, even, the infrared) region. The problem is
that there is no established theoretical procedure to calculate the parton densities in this region,
where many other more complicated multi-Pomeron graphs, not
accounted for in the BK-equation, become important. In particular,
the interactions between the gluons from two parallel `Pomeron-ladders'
are already not negligible at much lower HERA
energies \cite{BR}. Clearly the series alternates in sign.  The second enhanced correction
with two Pomeron loops gives a positive contribution, and so on. 
This is why the authors of Ref.~\cite{BM} wrote that
``we can not consider our
results as representing a reliable numerical final answer''. Moreover,
note that in \cite{BM}, just the first semi-enhanced Reggeon graph
was considered. It was demonstrated by Abramovsky \cite{Abrb,Abr} that the inclusion of
more complicated Reggeon diagrams may strongly diminish the effective
value of the triple-Pomeron vertex.  In particular, it is found that including graphs with one and
two extra Pomerons reduces the effective value of the triple-Pomeron vertex $V_{3P}$ by a factor
of 4 \cite{Abrb}.

\begin{figure}
\begin{center}
\includegraphics[height=3cm]{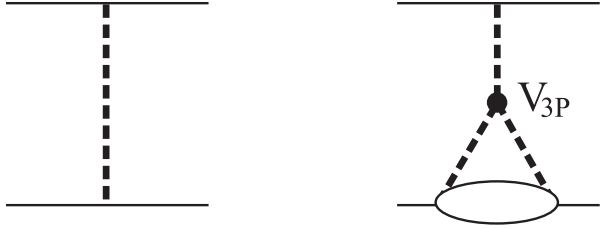}
\caption{The leading Reggeon contributions to the total and the single diffractive
dissociation cross sections.  The dashed lines correspond to Pomeron exchange.}
\label{fig:SD}
\end{center}
\end{figure}

 From the formal point of view, if we work perturbatively and include only the first Reggeon diagrams,
we can estimate the importance of the semi-enhanced correction by relating the ratio
of the contributions to exclusive Higgs production, 
$\sigma_H$(Fig.~\ref{fig:H}(d))/$\sigma_H$(Fig.~\ref{fig:H}(b)), to the ratio $\sigma_{\rm SD}/\sigma_{\rm tot}$.
Here $\sigma_{\rm tot}$ and $\sigma_{\rm SD}$ are the total and single diffractive dissociation cross sections
respectively, as computed from Fig.~\ref{fig:SD}.  We see that we have the ratios of equivalent Regge diagrams.
However, in the first ratio we need to include an AGK factor \cite{AGK} of 4; one factor of 2 since the Higgs boson may be emitted from either the left or right Pomeron in Fig.~\ref{fig:H}(d), and another factor of 2 as the cross section is given by the square of the amplitude.  Thus in terms of the simplest Regge
 diagrams we obtain 
\be
\frac{\sigma_H^{(d)}}{\sigma_H^{(b)}}~=~4~\frac{\sigma_{\rm SD}}{\sigma_{\rm tot}}
\left(\frac{{\rm ln}\sqrt {s/M^2_H}}{{\rm ln}(s/s_0)}\right)~\simeq~0.1
\ee
at the LHC, where the ratio in brackets is to allow for the different rapidity intervals available for the
triple-Pomeron vertex, $V_{3P}$.  The numerical evaluation of 0.1 is obtained using 
$\sigma_{\rm tot}\simeq 100$ mb \cite{KPTM,KMRsoft,GLMNP}, $\sigma_{\rm SD} \sim 10$ mb,\footnote{$\sigma_{\rm SD}$ is
observed\cite{Dino} to be
already practically independent of energy by $\sqrt{s} \simeq 500$ GeV.} ln$(s/M_H^2) \simeq 9$ and
ln$(s/s_0) \simeq 18$ for the LHC energy.  This estimate of the size of the semi-enhanced contribution is
much less than that given in \cite{BM}.  The arguments employed in this paragraph, and in \cite{BM}, are based
on perturbative estimates using the simplest Reggeon graphs, whose validity is questionable close to the saturation regime. The true parameter of the perturbative series is not the QCD coupling $\alpha_S$, but 
the probability of additional interactions, which however tends to 1 as the saturation region is approached.

Let us discuss this in more detail. Note that, starting from perturbative theory, we arrive in the
strong coupling regime.  The main contribution comes from the rescattering of partons with low $k_{t,4}<Q_s(x_4)$,
that is from the region where the probability of rescattering is of the order of 1.  So we must consider the
possibility of double counting.  Indeed the calculation of the ``soft'' survival factor, ${\hat S}^2$, 
in \cite{INC, KMRsoft}
used the phenomenological $pp$-amplitude obtained from fitting to ``soft'' data.  This amplitude, shown 
 by the left vertical line in Fig.~$\ref{fig:H}$(c), already includes the
enhanced Reggeon diagrams like that shown in Fig.~$\ref{fig:SD}$; that is it accounts for the rescattering of the whole proton wave function 
including all the intermediate and ``wee'' partons.  Thus we do not need to consider the contribution 
of Fig.~$\ref{fig:H}$(d), but instead the {\it difference} between the enhanced contributions to exclusive
Higgs production and the enhanced corrections hidden in the {\it phenomenological} soft $pp$-amplitude.
Here we may appeal to the Good-Walker approach \cite{GW}. Qualitatively, we expect that the component of the proton wave function, which contains the Higgs boson, will have smaller size and a smaller number of wee partons than
in a normal proton.  The probability of a soft rescattering for this component is, most probably, smaller, that is
the gap survival factor is larger, than that calculated using the ``experimental'' elastic $pp$-amplitude.
So, contrary to Ref.~\cite{BM}, we may find that the corrections from the enhanced diagrams could even
enlarge the predicted exclusive Higgs cross section.  However the probability of a soft rescattering is
mainly driven by the spatial distribution of the valence quarks, so we do not expect the effect to be large (see e.g. \cite{enh,SWILL}).

We conclude that there are theoretical and phenomenological reasons why the semi-enhanced
corrections are expected to be small at LHC energies, and will not appreciably affect the estimates, outlined in Section 2, obtained
for the cross section of the exclusive process $pp \ra p+H+p$.
Indeed, first, the correction comes from the `saturation' (or even the
infrared) region, where the global parton analyses which include the low $x$ HERA structure
function data, show that, at low $Q^2$, the
gluon density decreases as $x$ decreases below $10^{-3}-10^{-4}$.
Moreover, the concept ``gluon density $(f_g)$'' is not well defined in the saturation domain.
When we enter the strong coupling regime of saturation we have to rely more on phenomenological
arguments.  One of these is the `$\Delta Y$' threshold effect, which arises from the NLL correction
to the triple-Pomeron vertex; it is expected to strongly suppress the
correction when $x_H > 0.01$.
However there is a more direct way of checking the smallness of the semi-enhanced 
hard rescattering correction.  To this we now turn.

\section{Enhanced diagrams: experimental information}

There is a good way to experimentally probe the importance of the semi-enhanced rescattering
correction.  It is the observation of leading neutron production in
inelastic events at HERA, in which the neutron is measured with Feynman $x$ in the region $x_L\simeq 0.7 - 0.9$. This process, $\gamma p \ra Xn$, is mediated by pion exchange.
The gap corresponding to pion exchange may be filled by the
secondaries produced in the rescattering of intermediate partons,
in exact analogy with the case of exclusive Higgs production. Due to
the relatively large values of the momentum fraction ($1-x_L$) transferred
across the gap, here the rapidity interval available for the
triple-Pomeron vertex is already large enough at HERA energies. Since
 the whole correction, after the integration over the rapidity of the
triple-Pomeron vertex, is proportional to the available rapidity
 interval, which grows with the initial photon energy, one has to expect
that the probability to observe a leading neutron (that is to observe a gap) must fall down with
energy. However this is not observed experimentally.  The leading neutron data can be found
in \cite{zeus1}-\cite{h12}, and a detailed analysis and discussion of the data is given in \cite{LN}.
These HERA data show a flat dependence on the incoming photon energy; see, for example, Figs. 7 of \cite{h1},
and Tables 14, 18 and Figs. 11, 12 of \cite{zeus2} which show, for fixed $Q^2$, the same probability\footnote{That is
the same probability, ${\hat S}^2$, to observe the rapidity gap associated with pion-exchange.} to
observe a leading neutron for
values of $x_{Bj}$ which decrease by more than an order of magnitude corresponding to an increase of the
photon laboratory energy by more than a factor of 10. 
The flat behaviour
provides a strong phenomenological argument in favour of a {\it small}
semi-enhanced correction.

Soon there will be another way to check experimentally the role of the
semi-enhanced rescattering corrections.  That is from the measurements of exclusive $\gamma\gamma$ production,
$\pp \ra p+\gamma\gamma+\bar{p}$,
at the Tevatron, and subsequently at the LHC where large $\gamma\gamma$ masses should be accessible. Three candidate events have already been observed in Run II at the Tevatron \cite{gamgam}.
These hint at a cross section that is even larger than that predicted by a calculation \cite{KMRS}
based on a similar mechanism to that described in Section 2, that is without the semi-enhanced correction.
Recall that the estimate of the correction in \cite{BM} significantly reduces the size of the exclusive cross section.
These Tevatron data are preliminary, and we await definitive measurements over a range of masses of the
 $\gamma\gamma$ system.
In particular, if measurements of $\gamma\gamma$ production with $M=10-20$ GeV were available, it should be
possible to confirm the prediction for the exclusive production of a SM Higgs with
$M_H=120-140$ GeV to the order of $30-50\%$. Moreover, if we account for the NLO corrections to $gg \ra \gamma\gamma$ then the uncertainty could be reduced to $10-20\%.$

\section{Conclusions}

The prediction for the cross section of  central exclusive diffractive production of new heavy objects
at the LHC is very important.  In particular, the Higgs production process 
$pp \ra p+H+p$ offers many advantages for
experimentally probing the Higgs sector, and, indeed, in some regions of SUSY parameter space can even be the Higgs
discovery mode.  The expected Signal-to-Background ratio is promising, but the event
rate (at least for a Standard Model Higgs) is low.  It is therefore crucial to check the existing predictions.
One recent check was
carried out in Ref.~\cite{BM}.  In this paper the basic ingredients of the calculation outlined in Section 2
were confirmed.  However the authors of \cite{BM} went a step further.  Their aim was to quantify the possible importance of the so-called enhanced diagrams.  Indeed, they calculated these contributions perturbatively and came to the conclusion that they could be significant, and could reduce the predicted event rate, although they drew attention to
 the limited validity of perturbative
 procedure.  We therefore addressed this issue
in this Note.  In Section 3 we presented arguments which indicate that the enhanced corrections will be
small at LHC energies, and will not appreciably affect either the value, or the uncertainty, of the previous predictions.
One reason is that there is just not sufficient room in rapidity for the triple-Pomeron vertex.  The LHC
is a bit below threshold for this contribution to be important.
Then, in Section 4, we described how measurements of leading neutrons at HERA clearly confirm the
smallness of these enhanced corrections.

\section*{Acknowledgements}

We thank Aliosha Kaidalov for particularly valuable discussions on this topic. We also thank
J. Bartels, S. Bondarenko, K. Kutak and L. Motyka for informative discussions on their paper. MGR would like to thank the IPPP at the University of Durham for hospitality, and ADM thanks the Leverhulme Trust for an Emeritus Fellowship. This work was supported by the Royal Society,
the UK Particle Physics and Astronomy Research Council, by the grant RFBR
04-02-16073, and by the Federal Program of the Russian Ministry of Industry, Science and Technology SS-1124.2003.2.


\end{document}